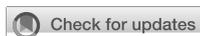

# Evaluating the Talbot-Plateau law

Ernest Greene* and Jack Morrison

Department of Psychology, Dornsife College of Letters, Arts and Sciences, University of Southern California, Los Angeles, CA, United States

The Talbot-Plateau law asserts that when the flux (light energy) of a flicker-fused stimulus equals the flux of a steady stimulus, they will appear equal in brightness. To be perceived as flicker-fused, the frequency of the flash sequence must be high enough that no flicker is perceived, i.e., it appears to be a steady stimulus. Generally, this law has been accepted as being true across all brightness levels, and across all combinations of flash duration and frequency that generate the matching flux level. Two experiments that were conducted to test the law found significant departures from its predictions, but these were small relative to the large range of flash intensities that were tested.

## Introduction

> When a certain place in the retina is stimulated always in the same way by regular periodic impulses of light, then, provided each recurrent stimulus is sufficiently short lived, the result is a continuous impression, equivalent to what would be produced if the light acting during each period were uniformly distributed over the entire time (Von Helmholtz, 1910, V2, p. 207).

In the quote above, Von Helmholtz (1910) is referring to a principle first articulated in the mid-1800s, which is now known as the Talbot-Plateau law. It is a rather remarkable claim about how brief, successive, light flashes will combine to determine the net perceived brightness of a given stimulus. As an initial requirement, the flash frequency must be sufficiently high to be perceived as steady/fused meaning that one does not perceive any flicker in the stimulus source. The Talbot-Plateau law asserts that the perceived brightness of a flicker-fused stimulus will be equal to the brightness of a steady stimulus when the average intensity of the two are equal. Stated differently, both stimuli must be delivering the same amount of light energy per unit time. Talbot's initial analysis began with the concept of visual persistence, which had been noted by Newton a century earlier (Newton, 1704; Talbot, 1834). Newton was discussing the streak of light that one can perceive from the motion of a glowing ember. He proposed that one could measure the duration of visible persistence by attaching the glowing ember to a spinning wheel and measuring the time it takes for the trailing streak to form a complete circle. Talbot continued the analysis as follows:

> The question deserves consideration whether the eye receives from this circular ring exactly the same quantity of light which it received from the much smaller surface of the coal at rest… If, then, the total quantity of light remains the same, it follows that its apparent intensity must have diminished in exactly the same proportion as its apparent area has been enlarged [spread around the perimeter] (Talbot, 1834, p. 328).

Talbot was asserting that the total light energy from an ember trace that has completed a full circle will be the same as the total energy from a steady ember across the same time interval. That energy is being spread across the black background of space, thus providing the streak. He further inferred that within a single period of a flickering light, the energy of the flash could fill the interval of darkness that follows.

The perceptual effect of light stimulation was a major topic of interest to physicists throughout the 1800s. One could spin a disk that had some sectors painted white and others painted black and note the speed of rotation at which the succession of black and white would blend and be perceived as solid gray. The relative areas of black and white would determine the shade of gray. Talbot's further inference was that the shade of gray would appear the same as a gray zone being illuminated by a steady source that provided the same quantity of light energy. Plateau (1830) had reached the same conclusion a few years before.

But this inference cried out for supporting evidence. Talbot provided this by comparing the perceived brightness of a disk that was illuminated with intermittent flashes to another disk providing steady light. Brightness of the steady source was controlled by two polarizing filters, these being wafers of calcite (spar) that came to be known as Nicols filters. The filters would transmit light when the polarizing orientations were aligned the same, and would block progressively more light as they were rotated into misalignment. Adjusting the relative orientation of the filters provided a way to quantify the amount of light being passed, and Talbot reported that the brightness of the flicker-fused stimulus matched the brightness of the steady source when the two delivered the same amount of light energy.

Talbot (1834) also briefly described experiments based on creating flash sequences with a spinning mirror that reflected a beam of sunlight. Intensity of the steady source was controlled by adjusting the aperture through which the beam passed. His experiments were suggestive, if not experimentally rigorous, but Stewart (1888) claimed support for Talbot's law with experiments that used the mirror method. For this, a light beam was bounced off a spinning mirror and by further reflection was carried into the eye of the observer. With each rotation a beam from the lamp would pass across the eye and be seen as a flash. The mirror revolved at a rate that assured fusion of the flashes, so the image appeared to be steady. For the steady comparison, the beam from a second source was reflected off a stationary mirror and then brought to a location that was adjacent to the flicker-fused stimulus. The intensity of the light was determined using the inverse square of distance. A given rotation speed would change the duration of flashes and simultaneously alter the frequency, providing compensation for total energy, as required by the Talbot-Plateau law. Stewart claimed to find support for that concept at all rotation speeds, even with flash durations as low as $1.2 \times 10^{-7}$ s (120 nanoseconds).

Fick (1863) said that the law did not hold when the stimulus was very bright or very dark, but most investigators during this period embraced the claim that the visual system was making precise assessment of light energy at all stimulus intensities. The Talbot-Plateau predictions were generally accepted as being true whether one was comparing bright or dim stimuli. It was generally accepted as being true whatever the ratio of light and dark intervals in the flash sequence, which is commonly known as "duty cycle." Von Helmholtz (1910) supported the concept, as reflected in the quote provided at the outset. He did so even though there were numerous reasons for doubting that it could be valid as an overarching principle. Von Helmholtz (1910), p. 180 noted that bright objects appear brighter in dim illumination, and cited examples of enhanced brightness when the frequency of stimulation was below the fusion threshold. He cited Purkinje that colors manifest differential brightness as a function of ambient illumination (Von Helmholtz, 1891), which he correctly attributed to the differential activation of rods and cones.

There were other findings that should have counseled caution about assuming a simple and precise mechanism for generating brightness perception. As one spins a disk painted with black and white sectors, at some disk speeds one sees brilliant colors, these being known as Benham or Fechner colors (see Bagley, 1902). Helmholtz dedicated a substantial amount of attention to the pulsed conditions that yield perception of colors. It was clear that the mechanisms for color and brightness were interconnected in some manner, and it was (and is) far from obvious how these interactions might bear on the production of brightness, *per se*.

In addition, Gestalt theorists rejected the concept that the perception of brightness was precisely tied to the quantity of light being delivered by the stimulus. A small sampling of the insights provided by the Gestalt School can be found in Ellis (1938), who translated papers by Fuchs, Benary, and Gelb. Fuch (pp 95–103) reviewed assimilation of colors, including the factors influencing the perception of white, citing Wundt as an early source of insight about brightness assimilation. He also mentions the presence of brightness-suppressing ghosts at the intersections of Hering squares (also commonly known as a Herman grid). Benary (pp 104–108) described brightness contrast that can be produced by placing the stimulus to be judged adjacent to various forms. Gelb (pp 196–209) provided examples of color and brightness constancy. Many of these phenomena had been noted by 19th century investigators, so there were already ample reasons to doubt that the perception of brightness would be based only on the quantity of light being received.

We could now add the late Gestalt contribution of Kanizsa (1979) who demonstrated that shapes implied by collinear lines and line terminations appear brighter than background. We see differential brightness of zones lying in shadow (Gilchrist, 1994), and the Craik-O'Brien-Cornsweet illusion continues to give pause (Cornsweet, 1970). All in all, it strains credulity that the visual system would be subject to so many sources of spatial and temporal influence and yet the mechanism for integrating successive flashes would precisely register the net light energy being delivered.

Notwithstanding the reasons for doubt, there were a number of claims of support from experiments done in the early 1900s. Sherrington (1902, 1904) tested the Talbot-Plateau principle using monoptic and dichoptic stimulation from beams passing through holes in rotating cylinders. Unfortunately, his description of method was insufficient to provide assurance that the brightness of the flicker-fused stimulus matched that provided by the steady light source.

Hyde (1906) used a rotating disk with open sectors that could chop a beam of light and thus produce flash sequences. He had special concern about the ability to quantify the amount of light being emitted from the source, given that the inverse square law is valid only if one has a point source. His illumination was from the element of a Nernst glow lamp with the diffusion globe removed, and extensive calculations were applied to correct for the fact that it was elongated and not a proper point source. He provided extensive data in support of the Talbot-Plateau law with various rotation speeds and open-sector ratios.

Beams (1935) provided a brief report claiming confirmation of the Talbot-Plateau law using a rotating-mirror for generating flash sequences and Nicols polarizing filters to control the intensity of the steady stimulus. He provided minimal details about experimental methods, and did not report any data other than saying that the principle was supported for flash durations ranging from $10^{-6}$ to $10^{-8}$.

Gilmer (1937) conducted two experiments, one that produced flash sequences with a rotating mirror and the other with a rotating disk. Intensity of the steady source was adjusted using Nicols filters. He reported support for the Talbot-Plateau law for flash durations ranging from $10^{-2}$ down to $8 \times 10^{-9}$ s. Brightness was evaluated across the range from 5 to 50 candles/m$^2$.

Bartley (1937, 1938a, 1939), equated the brightness of flicker-fused and steady stimuli, mostly to serve as a baseline for evaluating brightness enhancement at flash frequencies that were too low to produce fusion, i.e., the Brücke (1864) effect. Bartley (1938b) suggested that a "dark process" might be active during the "off" portions of a flash sequence. This could be seen as presaging a dual-channel hypothesis for encoding luminance, as will be discussed subsequently. However, he subsequently claimed that this signal would cease once the stimulus reached critical flicker frequency, which would rule out any contribution to the average intensity of a flicker-fused stimulus (Bartley, 1939).

The inventory of studies provided above likely has missed a few reports claiming support of the Talbot-Plateau law. The cited studies do, however, describe the research tools that would be available for testing the Talbot-Plateau law through the early part of the 20th century. We respect the ingenuity of early investigators in their efforts to evaluate the Talbot-Plateau predictions. However, for the range being planned for current work, those tools could not accurately measure the amount of light being delivered by the flicker-fused stimulus. To assess flash intensity using the inverse-square law would require positioning the source up to several hundred meters away. On an optical bench this could be done with mirrors, but the corresponding reduction of image size would not provide a usable comparison stimulus. Even if the Nicols filters were manufactured with the precision of modern polarizing filters, getting a 10% level of accuracy would require control over 0.01 degree increments of filter position. Modern, electronic equipment makes it possible to examine a wider range of conditions and more accurately control timing and intensity of the stimulus.

Szilagyi (1969) used modern electronic equipment to test what he described as the Bunsen-Roscoe law, though in fact he was evaluating the Talbot-Plateau principle. Three respondents adjusted current for an LED that was flashing at 30 Hz until it matched the brightness of a second LED that was providing steady emission. Current was converted to intensity according to separately calibrated current vs. intensity measurements. Current pulses that drove the emission were calibrated for linearity across about five orders of magnitude. Rise and fall times of the pulses were roughly 1 microsecond. He varied duration of current pulses (yielding flashes) from 1 microsecond to 10 milliseconds. The data from three respondents fit along a straight line on a log–log plot at a slope of −1, affirming the Talbot-Plateau predictions for these test conditions. However, the flicker-fused stimuli were displayed against only one level of steady light emission and ambient light level was not specified. At best this provides support for the Talbot-Plateau principle across a limited range of treatment combinations.

An earlier report from this laboratory included a limited test of the Talbot-Plateau principle with stimuli delivered from an LED array (Greene, 2015). An initial experiment established that a sequence of 1.3 microsecond flashes, delivered at 24 Hz, was above the flicker-fusion threshold. Then each trial of the following experiment presented letter patterns twice, as steady and flicker-fused stimuli, randomly ordered, each for 750 milliseconds. Flash intensity was varied across a range as departures from the Talbot-Plateau prediction, expecting that higher intensities would be seen as brighter than the steady display and lower intensities would appear darker. Respondents were asked to say which of the two displays was brighter. (Letter identification was not requested, but providing this stimulus diversity helped keep respondents engaged in the task). For those displays in which the steady and fused flicker displays were judged to be equally bright, the calculated average energy of the flash sequence was very close to the Talbot-Plateau prediction. The results provided some support for the Talbot-Plateau principle, comparable to the Szilagyi (1969) findings, but still insufficient for assessing its validity across several brightness levels and various flash duration/frequency combinations.

Here we report the findings from two experiments. The first varied flash duration of flicker-fused stimuli across five orders of magnitude against five octave levels of steady stimulus intensity. The second experiment displayed the same five flash durations to assess brightness judgments in a lower range of stimulus intensity. This extended the range across which flash intensity was tested to roughly seven orders of magnitude.

## Methods

### Stimulus display equipment

Experiments were executed using a 64×64 array of AlGaInP light-emitting diodes (LEDs), designated as the display board, which have peak emission at 630 nanometers (red). Diameters and center-to-center spacing of the LEDs were 5 mm and 9.4 mm, respectively, and the horizontal and vertical spans of the full array were each 60 cm. At the observation distance of 3.5 m, the visual angles formed by these spans are 4.92 arc´, 9.23 arc´, and 9.80 arc°.

A Mac G4 Cube running Tcl/tk custom applications under OS-X provided instructions to a Propox MMnet101 microcontroller. The microcontroller, running at 16 MHz, provided machine-language instructions to the stimulus display. The microcontroller crystal had a stability of 50 ppm, and the average speed for processing firmware instructions was 12 MIPs. This system allowed for nominal specification of treatment durations as short as 1 µs. (See the Supplementary material).

None of the experimental work involved color comparison, so it is appropriate to report the intensity in radiometric units–microwatts per solid angle (µW/sr). Further, physiological studies of photoreceptors with monochromatic or LED light sources often report stimulus energy in radiometric units (Schnapf et al., 1990; Packer et al., 1996; Schneeweis and Schnapf, 1999; Field et al., 2009; Cangiano et al., 2012; Cao et al., 2014). With such narrow-range light sources the responses of red and green cones as a function of intensity are very similar (Schnapf et al., 1990).

Intensity was measured using a Thorlabs PM100 radiometer with S120C calibrated silicon photodiode sensor. The calibration process started with power measurements of the display at 1 m, with a single LED turned on and taking readings across a range of LED voltage settings. The reading in microwatts (μW) was converted to radiant intensity in μW/sr by dividing the solid angle of the sensor (at 1 m, the sensor solid angle is essentially the same as its area in m$^2$). Then, power measurements were recorded over a wider range of voltages, at small increments, with the sensor placed directly against the display. These readings were scaled to match the 1 m intensities for corresponding voltages. This calibration produced a table of 100 samples from 0.0001 to 70,000 μW/sr. Experiment applications used linear interpolations from this table to convert a requested intensity to the necessary LED voltage.

Ambient illumination in the test room was 10 lux, as measured with a calibrated Tektronix J1811 photometer. At this light level, average pupil diameter was previously found to be 6.66 mm. This provides a basis for calculating the total energy of an LED flash, in photons, as seen by the respondent: radiant intensity (μW/sr) x flash duration (μs) x 9.019 (which adjusts for wavelength, pupil diameter, and source distance).

Voltage measurements were taken at an LED to confirm that the number of dots turned on at one time had minimal effect on actual intensity. Turning on 48 dots on a 64-dot module reduced the voltage less than 3% compared to turning on a single dot.

Oscilloscope traces were captured with an Advanced Photonix PDB-C156 PIN silicon photodiode in unbiased, unamplified photovoltaic mode. This is a fast photodiode that has a response time of 15 ns. The traces were taken with an appropriate load resistor to convert the current output into a voltage that was measured by a 1X voltage probe. Flash intensity was verified by comparing oscilloscope traces for flashing and steady emission.

Rise and fall times for this kind of LED are each 30 ns or less. For long flash durations this is nominal and the oscilloscope trace can be described as a square wave. For the 1 μs flash one sees a shoulder at the top of the rising phase and a belly on the falling phase, providing counterbalance of light emission.

All durations of flashes were also found to have a 0.3 μs transient of light emission at the offset of the voltage command, designated as an "off pulse," due to flyback voltage from LED inductance. The amplitude of the off pulse gets added to the overall quantity of light from the flash. To determine the total light energy being delivered, additional radiometer measurements were taken during periodic flashing at 500 Hz and higher (well above the meter analog averaging bandwidth of 30 Hz) for a range of flash durations and intensities. Traces did not show any differential in flash characteristics for single flashes versus those being driven at this frequency. These measurements were compared to steady intensity, and an empirical time-dependent compensation formula for the voltage-to-intensity conversion was determined.

For simplicity, flash durations used as experimental treatments are given as integers, and the reported intensities are the average of intensities across all portions of the flash, including the off pulse. Therefore duration x intensity provides a proper index of total flash energy, within the resolution of voltage command increments, irrespective of the flash duration that was commanded.

## Stimulus letters

Each letter of the alphabet was represented by a discrete pattern of dots, as illustrated in Figure 1. The letters were styled as Arial 33-point TrueType fonts. Supplementary Table S1 lists the number of dots in each letter as well as letter dimensions, specifically heights and widths in dot count and degrees of visual angle.

## Approval of protocols and informed consent

Experimental protocols were approved by the Institutional Review Board at USC. Sixteen respondents were recruited from the USC Department of Psychology Subject Pool, each having a broad choice among alternative experiments being conducted. Eight of the respondents provided data for the first experiment, and eight others provided data for the second experiment. Each respondent was provided with a general description of the display conditions and the judgments that were required. Each was informed that participation could be discontinued at any time without penalty.

## Task demands

There were several experimental conditions that were the same for both experiments. Treatment levels for the various experiments were developed through pilot testing of respondents. Unless noted otherwise, once treatment levels were adopted for a given experiment, the data from all subsequent respondents have been included in the statistical analysis and modeling.

For a given display, all the dots forming a given letter pattern were activated together, i.e., providing simultaneous emission from all the dots forming the pattern. For a given experiment, the presentation order for each combination of treatments was determined at random for each respondent. Respondents were tested individually. Each judged the stimulus displays using both eyes, allowing vision correction as needed. No feedback about the accuracy of the response was provided. The task did not require rapid response, but responses were usually provided with minimal delay. Responses were recorded by clicking on-screen buttons. Neither the respondent nor the experimenter was provided with any information about what treatment was being displayed on a given trial. Most test sessions were completed in about 40 min.

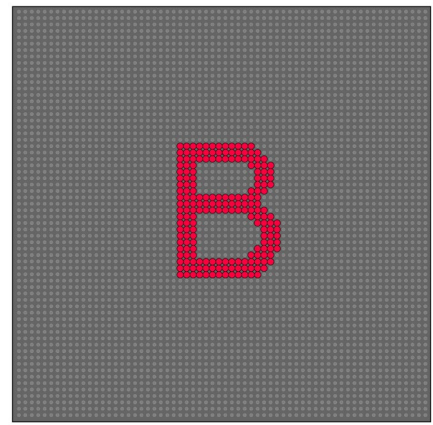

FIGURE 1
Letters were displayed as ultra-brief flashes or steady emission from a specific subset within the 64×64 array of LEDs. The background array represents the LEDs that did not emit any light. The pattern LEDs are enlarged in this illustration to reflect the increase in stimulus salience provided by light emission.

## Experimental treatments and protocols

Experiment 1 provided treatments to evaluate predictions of the Talbot-Plateau law across an extended range of flash duration and intensity levels. Letter patterns were displayed as stimuli, but respondents were not asked to identify the letters, only to judge the relative brightness when the patterns were shown with steady emission of light and also with a 24 Hz flash sequence, i.e., as a fused-flicker display. This experiment used an adaptive protocol, wherein the intensity of flashes was adjusted up if the fused-flicker display appeared dimmer than the steady display, and adjusted down if it appeared to be brighter. The steady and fused-flicker displays were each shown for 500 ms on a given trial, with a 300 ms interval between displays.

There were five steady intensity treatment levels, these being 0.01, 0.02, 0.04, 0.08, and 0.16 μW/sr. Flash durations of 1, 10, 100, 1,000, and 10,000 μs were used for the fused-flicker displays, wherein the intensity to be used with a given flash duration was set near the level predicted by the Talbot-Plateau law, these being subject to adjustment on the basis of respondent's assessment of relative brightness. The five steady intensity levels combined against five durations (with corresponding flash intensities), which provided 25 treatment combinations. The letter to be used for a given treatment combination was chosen at random.

For the initial display of a given treatment combination, the intensity of the flash was set at the intensity that would be predicted by the Talbot-Plateau law, but with a 20% adjustment, randomly making the flash brighter or dimmer than the prediction. Whether the steady or fused-flicker display came first or second on a given trial was determined at random.

The respondent was required to judge which of the two displays was brighter, the first (scored as 1), the second (scored as 2), or they appeared equal in brightness, i.e., Same (scored as 0). Once the experimenter recorded the verbal response of the respondent, the computer took note of which stimulus had come first, and determined whether to increase or decrease the intensity of the flash sequence on the next trial that treatment combination was displayed. Increments or decrements of flash intensity were in steps of 15% from the value delivered on a given trial.

On initial trials a few respondents were only able to see one pattern rather than two. Pilot work had determined that respondents were always able to see a steady display that was shown at one of the five treatment levels. However, the initial adjustment of the fused-flicker stimulus could drop the flash intensity to a level where the pattern had insufficient energy to reach threshold, so the respondent was seeing only display of the steady stimulus. To deal with this possibility, respondents were told to report if they only saw one display, not two. This was scored as 3 by the experimenter, which automatically resulted in an increment of subsequent displays -- the same as if they had judged the fused-flicker display as being dimmer than the steady display.

A termination condition for the adaptive protocol was provided. For a given treatment combination, if the fused-flicker and steady displays were judged as being equal in brightness (Same) on five separate trials, no additional trials were run for that treatment combination. The total number of trials was capped at 400. Upon completion of testing, the responses were tallied for each treatment combination, the recorded intensities that had received a Same judgment were averaged, and those values are described as the "observed" flash intensities for purposes of evaluating the Talbot-Plateau predictions.

Experiment 2 tested whether the Talbot-Plateau law would extend to a lower range of flash and steady intensities, these being closer to perceptual threshold. Four levels of steady intensity were tested, specifically: 0.0024, 0.0048, 0.0072, and 0.0096 μW/sr. As in Experiment 1, flash durations were 1, 10, 100, 1,000, and 10,000 μs, each being used in a 24 Hz fused-flicker sequence at an initial intensity that was 10% above or below the Talbot-Plateau prediction. The total number of trials was set at 360. Other than the differences described above, all other experimental conditions were the same as for Experiment 1.

## Results

### Evaluating Talbot-Plateau law across more than five orders of magnitude

Experiment 1 asked respondents to judge letter patterns that were displayed twice on a given trial, once as a 24 Hz flash sequence and also as a steady display, both being shown for half a second. The anchor for brightness judgments was provided by the intensities used for the steady displays. Five octaves of steady intensity were tested, the dimmest being 0.01 μW/sr and the brightest being 0.16 μW/sr.

The 24 Hz flash sequences will also be described as flicker-fused displays, this serving to affirm that they appeared to observers as providing steady emission of light. Prior research had found the 50% threshold for flicker fusion to be about 15 Hz with 1 μs flashes. At 24 Hz, very few respondents reported seeing any flicker, and those that did saw it on only a few trials (Greene, 2015).

The display system provided control of flash durations and stimulus intervals with a resolution of 1 μs (see Methods for details). The flicker-fused treatments varied flash duration across five orders of magnitude, with the briefest flashes being 1 μs and the longest being 10,000 μs. Intensity of flicker-fused displays was varied using an adaptive protocol to determine the flash intensity that would appear equal in brightness to its steady-emission partner. With a given treatment combination, if respondents judged the flicker-fused display to be dimmer than the steady display, the flash intensity was raised on the next display of that treatment combination. The flash intensity was reduced if it was judged to be brighter. The values being reported and statistically evaluated here are the means across all judgments where the respondents said that the two members of the pair appeared to be equal in brightness.

Initial data plots of these judgments are shown in Figure 2, wherein the Talbot-Plateau principle has been reframed so that the data can be shown across the full range of treatment combinations as a single plot-line. In each panel of Figure 2, the observed matching flash intensity (log scaled) is shown on the ordinate and the value that would be predicted by application of the Talbot-Plateau principle is shown on the abscissa. The data for each of the eight respondents is provided in separate panels, with the group means being shown in the ninth panel. The plot-line in each of the panels reflects the Talbot-Plateau prediction, i.e., the observations would exactly match prediction if they fell exactly on this line.

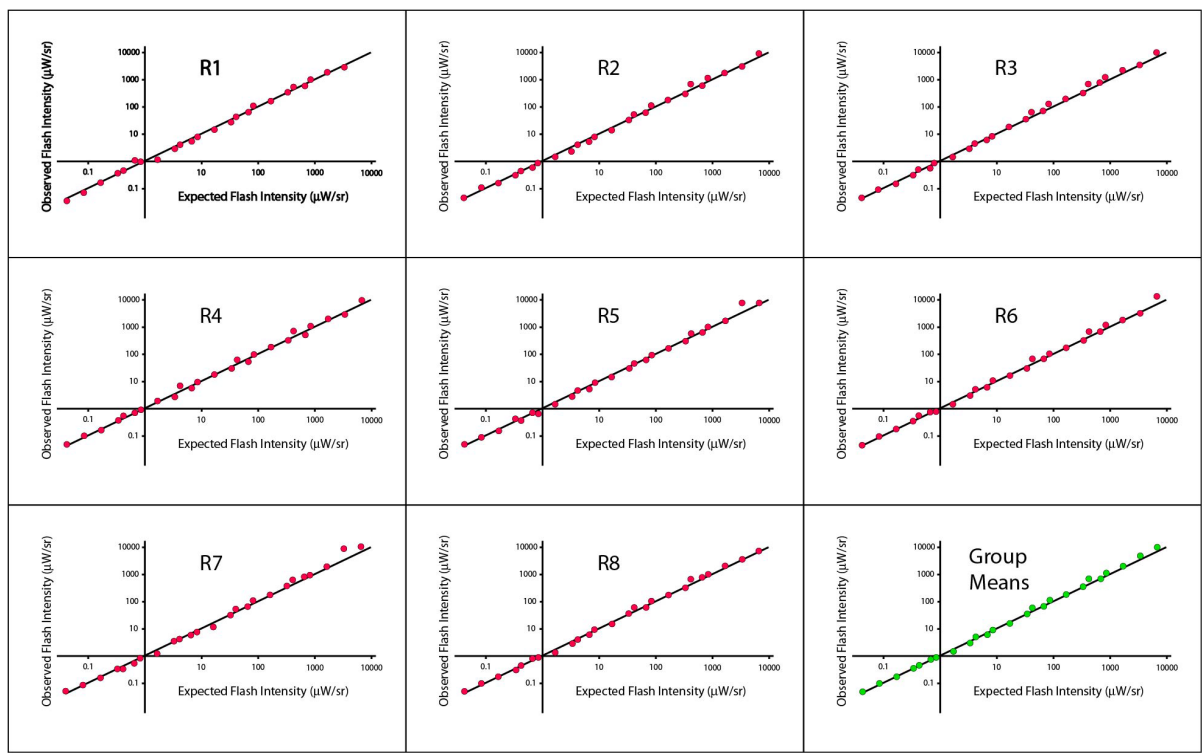

FIGURE 2
Observed flash intensities against expected flash intensities are plotted for each of the eight respondents of Experiment 1, with the ninth panel showing the group means. The expected flash intensities were derived from the Talbot-Plateau prediction for what flash intensities would match the steady displays at each of the treatment combinations.

Regression analysis for the group means provided an $R^2$ value of 0.9966, and a $p < 0.0001$. A mixed model was also applied to account for multiple points being from the same respondent, and the results were comparable. Regressions that were run on each respondent yielded $R^2$ values that ranged from 0.9962 to 0.9981, all significant at $p < 0.0001$. Pearson and Spearman correlations were both 0.9983, across respondents, the smallest Pierson and Spearman correlations were 0.9981 and 0.9977, respectively.

Figure 2 appears to provide strong support for the Talbot-Plateau law across an extended range of flash durations and intensities. One should remember, however, that the data are being shown on a log scale, which can shrink departures from prediction to the point that they appear insignificant. Differentials in intensity, when specified with less scale contraction, do not seem inconsequential. In particular, one can specify the ratio of observed to expected flash intensities, non-transformed, and find that they range in size from 0.85 to 1.55, with Talbot-Plateau prediction being a ratio of 1.00. So the flash intensity can be roughly 50% higher than the steady intensity at the point where the respondent judges them to be equal in brightness. There are conditions where a 50% increase in intensity can have a significant influence on perception, so it may be too soon to declare a ratio of 1.55 as being inconsequential.

With careful inspection of plot points across the panels of Figure 2, one may perceive small consistencies in placement relative to the plot-line, suggesting some micro-structure across the range. As evidence that this is not purely imagined, one can note that the deviations seen in the group means are about as large as those seen in the individual respondent data. This would not be the case if the respondent deviations were random departures from the Talbot-Plateau prediction. If that were true, the standard-error of the mean should be reduced, and the plot-points for the group should fall on the line or much closer to it.

Sources of consistent deviation from prediction can be more readily seen where the same data are shown using the original formulation of the Talbot-Plateau law, namely that the average intensity of the flashed sequence is equal to the intensity of the steady display. One should recall that the average flash intensity is the value derived from calculating the intensity of the flash multiplied by the duty cycle for a given flash period. Said otherwise, one is averaging the intensity of the flash with the dark portion of the flash period. According to the Talbot Plateau law, the average flash intensity should equal the steady intensity irrespective of the flash duration, for the duty cycle accounts for the differentials in flash duration. Therefore, all the flash duration plot-points for a given steady intensity should fall at the same location on the Talbot-Plateau plot line. One can see in Figure 3 that the observations did not quite match the predictions.

These data were not normally distributed, so differentials of treatment effect were tested with a nonparametric one-way ANOVA. This analysis found significant differences in mean values (Kruskal-Wallis test, $p<0.0001$) and variance measures (Ansari-Bradley one-way analysis, $p=0.0135$).

To examine for differentials in flash duration effect, a signed-rank test was applied to determine whether the average flash intensity minus the steady intensity differed significantly from 0. Median differences for 1 μs and 10 μs flash durations were 0.0061 and 0.0029 ($p<0.0001$ and $p=0.0437$), respectively, reflecting intensity settings that were higher than predicted by Talbot-Plateau. The 1,000 μs flash duration produced a median difference of −0.0060, this being lower than predicted and significant at $p<0.0001$. The 100 μs and 10,000 μs treatment levels had median differences of −0.0007 and 0.0011 ($p=0.0627$ and 0.4192), respectively, neither proving to be significantly different from the predicted zero value.

## Evaluating the Talbot-Plateau law across a low-intensity range

Experiment 1 examined the Talbot-Plateau law across a large range of intensities, these being at levels where brightness would commonly be evaluated. A second experiment extended the tests into a lower range that would be used to examine probability of recognition and/or detection as a function of flash intensity. Any departure from the Talbot-Plateau prediction could indicate a breakdown of physiological mechanism or a limit on the validity of the calibration process. Knowing how well the brightness judgments continued to match the Talbot-Plateau predictions could be useful in designing experiments that probed the threshold for shape recognition.

Four levels of steady intensity were evaluated, ranging from 0.0024 μW/sr to 0.0096 (so ending just below the lowest intensity of 0.01 μW/sr used in Experiment 1). The same five levels of flash duration were used, the minimum being 1 μs and the longest being 10,000 μs. At the highest steady display the predicted flash intensity at each duration was just under the lowest value in Experiment 1. For example, for the 1 μs flash sequence the lowest in Experiment 1 was 417 μW/sr and the highest intensity in Experiment 2 was 400 μW/sr. The lowest steady and flash intensities in this series were likely just above threshold. Whereas all of the respondents for which data is reported were able to see each display pair, one individual was unable to see all the displays, reporting either that only one pattern was shown or that none could be seen.

Figure 4 shows plots of means for individual respondents and for the group, with the plot-lines representing the Talbot-Plateau prediction. The plot-points for each respondent fall very close to this prediction across most of the range, with departures being most conspicuous at the low end of the range.

A log-scale regression was calculated across respondents. The analysis yielded an $R^2$ of 0.9950 and significance level of $p<0.0001$. Pearson and Spearman correlations were 0.9975 and 0.9978, respectively. A mixed model was also calculated to account for multiple points being provided by a given respondent, and the results were substantially the same. Regression models for each individual respondent provided $R^2$ values that ranged from 0.9902 to 0.9983, each having a $p<0.0001$, Across respondents, the smallest Pearson and Spearman correlations were 0.9951 and 0.9970, respectively.

Figure 5 plots the data from Experiment 2 rendered as average flash intensity. Here the departures from Talbot-Plateau prediction are more conspicuous than was found in Experiment 1. However, the range of intensities used in this experiment were much smaller, allowing data to be plotted on a linear rather than logarithmic scale. This could magnify the apparent size of departures from prediction.

Data were normally distributed, so a parametric one-way ANOVA was applied to examine the influence of flash duration on these brightness judgments. The mean differences were significant with $p<0.0001$. Each duration treatment was then examined with t-tests, to determine whether average flash intensity minus steady intensity differed from zero. Mean differences for the 1 μs flashes were brighter than predicted by the Talbot-Plateau law. The mean difference was 0.0012, which was significant at $p<0.0001$. The mean differences were smaller than predicted for 1,000 μs and 10,000 μs flashes, these values being −0.0010 and −0.0014, respectively. Each of the differences proved to be significant at $p<0.0001$. Treatment effects for 10 μs and 100 μs flashes were not significant ($p=0.0682$ and 0.3484, respectively).

We might speculate about the basis for significant deviation from the Talbot-Plateau prediction. It is possible that we have reached the limits of precision for the display equipment. If so, additional calibration efforts might provide future experiments with results that reduce or eliminate these departures. Alternatively, perhaps we are reaching the physiological limits for which the Talbot-Plateau law can apply. Flashes might become progressively less effective as the duration is reduced, as suggested by the ordered size of departure that can be seen in the Figure 5 plot. If so, one would expect the differentials to be larger as one approached the threshold of perception, but the scatter in Figure 5 is larger for the brighter flashes. Further, if one were reaching physiological limits for the flicker-fused mechanisms, one would expect the departures to be consistently to one side of the predicted plot-line, not equally balanced on each side of the line. We do not think the current findings provide a basis for setting limits on the validity of the Talbot-Plateau law.

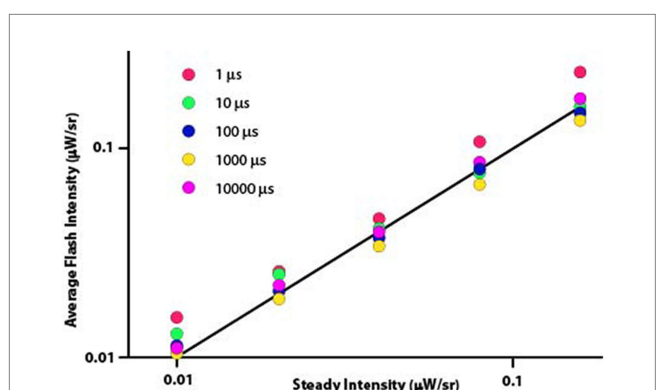

FIGURE 3
The straight line specifies the Talbot-Plateau prediction that average flash intensity will equal steady intensity when the two displays are judged as appearing equally bright. One can see systematic differences produced by flash durations, with the 1μs flashes diverging especially from the prediction.

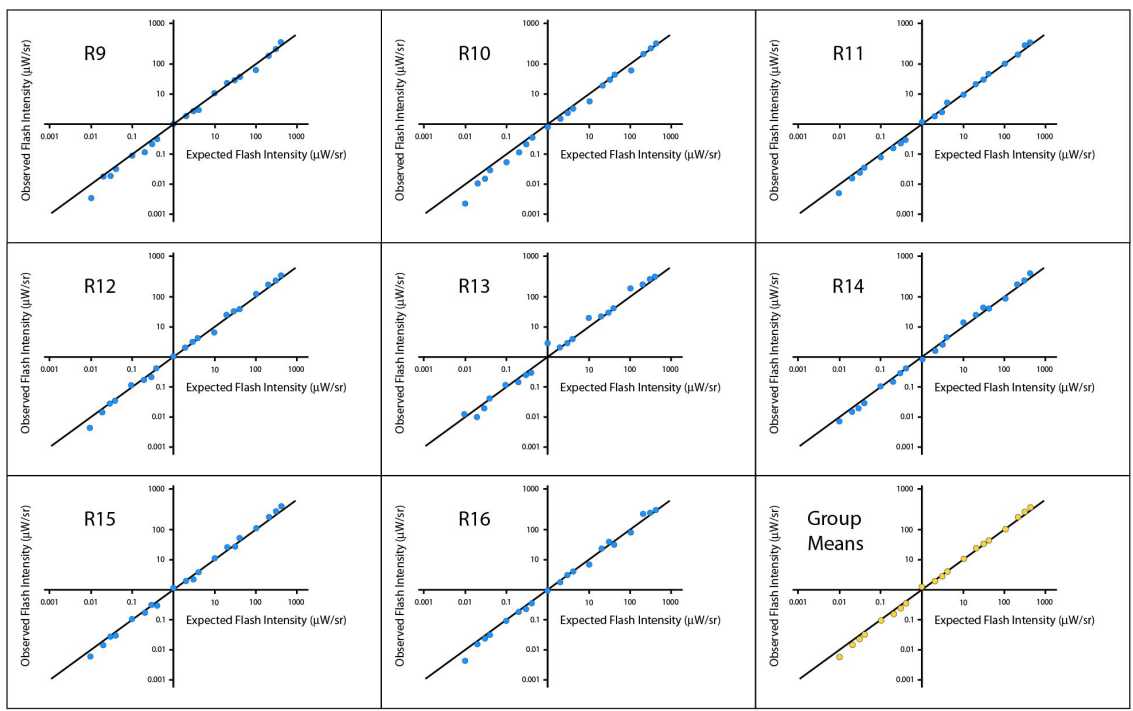

FIGURE 4
Experiment 2 tested the Talbot-Plateau predictions within low ranges of flash and steady intensities. The observed flash intensities, when plotted on a log scale for each of the respondents, were very close to the Talbot-Plateau prediction, this being the solid diagonal line in each of the panels.

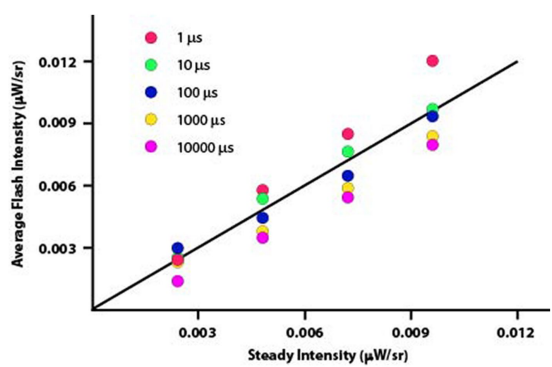

FIGURE 5
The data from Experiment 2 have been plotted as average flash intensity versus steady intensity. The Talbot-Plateau prediction, shown by the diagonal plot-line, is that the two displays will appear equal in brightness when the average flash intensity equals the steady intensity. One can see consistent departures from this prediction, with progressively greater intensity being required as the duration of the flashes was reduced.

## Discussion

### Generating a precise flicker-fused stimulus

The Talbot-Plateau law specifies that the brightness perception that is generated by the flicker-fused stimulus must match what is generated by a steady stimulus. This requires that the product of flash duration and flash frequency must be reciprocal, providing a constant level of light flux that exactly equals the flux of a steady stimulus. To explain this reciprocity, Ferree and Rand (1915) proposed that flashes produced progressively less impact as their duration was reduced. Thus a white or colored zone that would have a high level of brightness when not moving would be seen as darker if the spinning wheel provided only brief glimpses. The brevity of stimulation was viewed as providing less "intensity" of stimulation, so as speed of spinning increased the frequency of the flashes, the intensity being delivered by the white sector would be correspondingly reduced. While they made no mention of the Bunsen and Roscoe (1855) law, their proposal had much in common with its principles.

The photochemical mechanisms of the retina might well provide for reciprocity of duration and frequency, as specified by the Bunsen and Roscoe (1855) law. This law has been invoked for explaining duration/frequency reciprocity for threshold perception of short flashes, i.e., the Bloch effect (Bloch, 1885; Hartline, 1928, 1934; Brindley, 1952). Hartline (1928) examined the photoreceptor response in a number of arthropods, varying flash intensities and durations, and confirmed that the size of the response was determined by the total energy in the stimulus. In 1934 he again examined the issue using a rotating disk to control flash duration and frequency and reported reciprocity of light energy for flash durations of up to 100 milliseconds (Hartline, 1934). Szilagyi (1969) invoked the Bunsen-Roscoe law as the basis for finding equal brightness for flicker-fused and steady stimuli.

Even if the Bunsen-Roscoe law is correct, with photopigments registering every photon that the stimulus delivers, that activation must then be reliably conveyed to the brain where brightness is perceived. We need to specify the neural mechanisms that could accomplish the brightness matching predicted by the Talbot-Plateau law. Perhaps the light energy of the flash fills in the dark interval through visible persistence. The concept of persistence, sometimes characterized as afterimages, was central to the thinking of 19th century investigators. Von Helmholtz (1910), p. 212–215 commented on the weakening of persistent traces as ambient light levels increased, with a presumption that residual influence would cease to be present, or at least become imperceptible, in full daylight. Early discussions noted changes in the frequency at which fusion would occur, though specific measures were not reported until well into the 20th century. The early investigators readily embraced the idea that the brief exposure to light from a white sector of a spinning disk would provide sustained retinal activation that carried across the absence of light from the dark sector. Helmholtz, among others, gave substantial attention not only to temporal integration of black and white, but also to the blending of colors. One could readily assume there must be a limit to the temporal resolving power of the visual system that caps one's ability to distinguish the individual flashes. Once the frequency passed the resolution threshold, the flashes would fuse and be seen as a steady source of light. Walker et al. (1943) said that retinal persistence was surely responsible for flicker fusion, the mechanism being the same as in motion pictures.

The persistence concept might seem reasonable where light and dark intervals were of similar duration, but it becomes problematic for low duty-cycle displays where each flash is extremely bright and each dark interval is extremely long. For example, where the flash duration is one microsecond and flash frequency is 25 Hz, the dark interval that follows the flash is essentially 40 milliseconds. The intensity of the flash would be roughly 40,000 times brighter than the steady stimulus, which would have to be precisely measured and then drained at a rate that completely filled the interval without leaving any residual. An electrical engineer might describe the mechanism as being like a "leaky capacitor," where the capacitor is quickly charged by the flash energy and the voltage is drained off slowly at a steady rate. But the neurons would also have to precisely measure the length of the period between flashes and then adjust the rate of energy discharge according to how much energy the flash had provided. While it is possible that retinal circuits accomplish this task, we have no evidence at present that they have this ability.

One might also consider that the after-effects of stimulation can produce reverse perceptual effects, e.g., negative after-images. The energy provided by a flash might be eliciting a biphasic response, and there is a long history of work showing that the cessation of light generates its own signal. Let us now consider the possibility that the Talbot-Plateau principle is based on a system that counterbalances light and dark signals.

## Counterbalance of on and off retinal channels

Adrian and Matthews (1927), Granit (1933), and Hartline (1938) described OFF responses from retina that were generated by the cessation of steady illumination. Bartley (1939) asserted that the "dark phase" served as an "entity" that could suppress activation that was produced by the flash that preceded it. He did not specify what neural mechanism might be providing this counterbalance, but his formulations did imply some kind of signal generation. Many hundreds of articles have subsequently confirmed and provided details with respect to the anatomy and physiology of ON and OFF retinal mechanisms. Excellent reviews were provided by Schiller (1992), Schiller (2010), and Chalupa and Günhan (2004).

Enroth (1952, 1953) provided the first documentation that activities of ON and OFF retinal ganglion cells were relevant to flicker fusion. She recorded the frequency at which the cells in cat retina ceased to follow stimulus flicker, and described the cessation of activity as the flicker became fused. She asserted that flicker-fusion frequencies for the ON and OFF retinal ganglion cells matched what Dodt (1951) had reported for fusion of ERG pulses and subjective fusion in humans. This finding suggested that alternate activation of ON and OFF ganglion cells mediated perception of flicker, such that one no longer perceived flicker once the ganglion cells reached the limit of frequency that they could follow. However, specifying the frequency at which the flicker subjectively fuses, i.e., is no longer perceived as separate pulses of light, did not explain why the average intensity of a flicker-fused stimuli would equal the intensity of a steady stimulus. Further, defining cessation of flash "following" as the condition that would produce subjective fusion does not specify the nature of flicker-fused neural activity, or in particular, what the ganglion cells are doing in response to steady and flicker-fused stimuli. The mechanism for signaling the net energy of each stimulus was not articulated.

A number of investigators suggested that transient-responding channels change to provide sustained response once the frequency of flashes becomes sufficiently high (Granit, 1955; Arduini and Pinneo, 1962; Arduini, 1963; Pinneo and Heath, 1967). DeValois et al. (1962) observed differentials for ON and OFF channels using extracellular recordings from lateral geniculate nucleus of macaques. Firing rates either increased or decreased, roughly proportional to log intensity of the luminance. Arduini (1963) noted that above about 35 Hz the discharge of ON and OFF channels is constant at a level that is comparable to the average level of discharge being registered in a gross electrode. Pinneo and Heath (1967) proposed a two-stage process for registering flash sequences, with a detailed discussion of phasic responses transitioning to a frequency-fused tonic state. They also presented results of recordings taken from two patients who had depth electrodes implanted for therapeutic purposes into the optic tract and lateral geniculate nucleus. Neither patient had visual impairments. They used a steady light set at 2.5 lux, and varied flash frequencies from 1 to 50 Hz. Patients reported whether they saw flicker or steady light levels. With onset of flicker, net activity measured from the optic tract of the first patient increased with each increase in frequency up to a peak at 12 Hz. Above 12 Hz the frequency decreased with each step up to about 33 Hz at which point the activity level became constant for all higher frequencies. With stimulation frequencies of 33 Hz and beyond the flash sequence was perceived as being fused, with a brightness that was the same as the steady light. Optic nerve spiking was constant at this frequency and remained so at higher frequencies, and perceived brightness was unchanged. The second patient gave comparable results.

These findings support the hypothesis that ON and OFF channels work together to provide brightness perception, delivering brief (phasic) bursts of firing in response to sequential flashes, sustained (tonic) levels of firing when the frequency of the sequence becomes sufficiently high, and also providing sustained firing in response to a steady stimulus. If valid, one should see tonic activity being delivered from the retina as a function of light level. Barlow and Levick (1969) did locate some ganglion cells in the cat retina that manifested "unusually regular maintained discharge," the rate of firing being a function of the level of the adapting stimulus. These cells had ON or OFF concentric receptive fields, but they provided a very sluggish response to moving stimuli and lacked the brisk activation that was typical for other ganglion cells. They found only three of these cells, which they described as "luminance units." While such low density might seem adequate for registering the brightness of large zones, e.g., a light-blue cloudless sky, these would not seem to be suitable candidates for specifying the brightness of a small spot of steady light, or flicker-fused flashes from that spot.

Output of retinal activity is provided by optic nerve fibers, these being the axons of retinal ganglion cells. Gouras (1968) observed that some retinal ganglion cells of macaque responded to light increments with transient activity, while others gave tonic responses. Margolis and Detwiler (2007) examined tonic activity of ON and OFF ganglion cells, and report that OFF activity sustained by intrinsic pacemaker mechanisms. Zhou et al. (2017) provide an extensive evaluation of tonic and transient responses of retinal ganglion cells, outlining seven alternative mechanisms for how the two types of activation are generated.

Tikidji-Hamburyan et al. (2015) did multichannel recordings from ganglion cells in mice, providing full-field luminance across a 5 log range. A majority of the neurons registered luminance transitions irrespective of level, but a given cell provided stable responses at a specific steady luminance level. They found many instances where transient OFF activity could be elicited from ON cells, and vice versa. These results raise the intriguing possibility that the ON and OFF pathways register temporal contrast as a contribution to scene analysis (see below), but also provide steady activity that determines the brightness of a given region.

## Counterbalance of cortical mechanisms

Various researchers have debated the question of whether perception of brightness is based on integrating the flash responses in the retina or in cortex. Sherrington (1902, 1904) reported that stimulation of one eye yielded results that were consistent with the Talbot-Plateau principle, but dichoptic conditions that required synthesis of the stimuli in visual cortex did not. He reported that where the fused-flicker stimuli provided each eye with a different brightness, the cortex provided perception that was the average of the two. From this he concluded that Talbot-Plateau integration was accomplished at the level of the retina. Crozier and Wolf (1941) re-evaluated the issue and supported Sherrington's position. Shevell et al. (1992) have taken the opposite view, providing evidence for a cortical basis for contrast and assimilation of brightness using dichoptic stimulation. However, their focus was on center/surround receptive field mechanisms, so their results may not directly bear on the issue of Talbot-Plateau integration.

Bartley (1938a,b, 1937) suggested that cortex is responsible for assessing the brightness of flicker-fused stimuli. However, most of his rambling analysis was devoted to the Brücke (1864) effect--brightness enhancement--that is produced at frequencies that are below the fusion threshold. His work some decades later (Nelson and Bartley, 1964) provided a more explicit hypothesis which they described as an "alternation-of-response" mechanism. It proposed that brightness judgment is based on the sum of neural activity received in the visual cortex, noting that each input channel has a refractory period. A brief flash can elicit synchronized activity that produces a strong response. Various channel elements have different refractory periods, so a steady response produces non-synchronized firing that is at a lower net energy. Here again, it should be said that much of their work was directed toward evaluating Brücke enhancement of brightness at sub-fusion frequencies, and they gave very few details about experimental method. They provided minimal evidence in support of the claim that Talbot-Plateau integration is provided by the cortex.

During this period, extracellular recordings of neurons in cortex were failing to see steady, sustained responses that might reflect the perception of stimulus brightness (Hubel, 1958; Jung, 1964; Wurtz, 1969; Poggio, 1972). It was something of a surprise, therefore, when Bartlet and Doty (1974) reported the existence of "luxotonic" units in primary visual cortex of squirrel monkey, with similar findings subsequently reported for macaques (Kayama et al., 1979). Nearly half of the units studied by Bartlet and Doty (1974) and a quarter of those studied by Kayama et al. (1979) provided sustained response to diffuse light levels, while still manifesting both ON and OFF response properties. These neurons delivered a transient response with initial onset of the light, which provided the basis for the ON and OFF classification. However, after this initial transient activity, they continued with a steady firing rate for as long as the illumination remained the same. Rate of discharge was generally monotonic, either rising or declining as a function of light intensity over a range of at least three log units. Many of the cells also responded to other stimulus attributes, e.g., motion, providing for what might be called multiplexed registration of image information. Responses did not show any transitions that might reflect rod saturation as light levels were increased, supporting their hypothesis that the luxotonic units were being driven by cones, the influence being passed up through midget and parvocellular channels.

The proposal that steady brightness information might be signaled by tonic activity in two sub-populations, i.e., being driven by OFF as well as ON channels, was a departure from the concept that brightness would be specified by the firing rate of a single "luminance" population. Doty (1977) suggested the terms "photergic" and "scotergic" to describe the two classes of neurons, one that raises its firing rate in response to increasing luminance and the other doing so as the light level is reduced.

Doty and associates (Bartlet and Doty, 1974; Doty, 1977; Kayama et al., 1979) thought it likely that prior investigators had not seen steady discharge to steady light levels because a majority of the work had been done under anesthesia or using other pharmacological agents. They noted that anesthetics, even ones as mild as nitrous oxide, eliminated luxotonic responding in ON as well as OFF units. The phasic responses were not subject to this impairment, or at least the responses were less fully suppressed.

Peng and Van Essen (2005) provided further evidence supporting the dual-channel hypothesis. They monitored responses in V1 and V2 of macaque to a large zone that slowly oscillated across a large range of luminance values. Some neurons increased their firing rate as luminance increased and others increased their firing rate as luminance decreased, consistent with the findings of Doty and associates (*op. cit.*). Most of the neurons appeared to be tuned to a limited range of luminance. This would support a proposal that specific brightness judgments, i.e., perceived gray level, is provided by selective neuronal response to stimulus attributes, similar to orientation selectivity.

Smith et al. (2015) detail the segregation of ON and OFF channels in carnivores and primates as the luminance information passes from retina, through lateral geniculate nucleus, with convergence onto the layer 4 neurons in V1. The activity relays from there to the complex cells of layer 2/3, where the cells have overlapping ON and OFF response fields. These investigators found that the neurons in layer 2/3 responded to rising or falling luminance with firing rates that matched Doty's dual-channel hypothesis. Most of the neurons provided transient responses, but a small portion were thought to be "luxotonic" cells.

Yang et al. (2022) used multichannel extracellular electrodes to register activity from various layers of V1 cortex in alert monkeys. Superficial layers that receive the sensory input were more strongly activated by luminance of the stimulus surface than by the edges. However, edges dominated activity in the output layers of V1, and they suggest that the edge information contributes to "filling in" the perception of luminance within the interior of a stimulus.

## What evolutionary benefit?

A critical question remains, which is to explain what might be the functional benefit of a balanced dual-channel system. The historical claims for reciprocity of light and dark in the perception of brightness, formalized in the Talbot-Plateau law, appear to be valid across a very large range of light intensity, as well as diverse combinations of flash duration and frequency. The visual system is providing fairly precise encoding of the light and dark transitions being delivered by a flicker-fused stimulus. Such precision is not compelled by physical or chemical principles. One might invoke the Bunsen and Roscoe (1855) law that specifies reciprocity of flash intensity and duration in producing photochemical reaction products. The chemical cascade that is launched when the photopigment captures light might well conform to the Bunsen-Roscoe law. But from that point on we are dealing with physiological signal-generating mechanisms, and we do not think a high precision dual-channel system would evolve unless it had a functional benefit.

We submit that the balanced dual-channel mechanism is designed to register the luminance of each local zone in the image, providing stability of the perception in the presence of eye and stimulus motion. The eyes are in continuous motion as a scene is being scanned (Kowler, 2011; Gegenfurtner, 2016). Even when one is engaged in steady fixation of a target, the eye is quivering, drifting, and executing microsaccades. At one moment an object will be displayed to one patch of photoreceptors, and a moment later its image falls on a different patch. The same translation of image content would occur with stimulus motion. Precise recording of the luminance would contribute to translation invariance when registering a moving object or where eye movement changes image location.

Further, luminance may need to be registered with precision to correctly assess gradients of brightness. Those gradients can be critical for perceiving shape from shading, as provided by curvature of a surface (Kunsberg and Zucker, 2014; Antensteiner et al., 2018; Bartal et al., 2018; Holtmann-Rice et al., 2018). A consistent gradient of curvature would need to be reported as the object moved from one location in the visual field to another. If each zone in the retina were not precisely tuned for registering luminance, the shape of the surface would undergo constant deformation as the object moved or as the eye slowly scanned across it.

If eye motion is stabilized to prevent translation of an image, the luminance, contrast, and chromatic differentials can disappear rapidly (Yarbus, 1967; Murakami et al., 2006). Similar changes take place if the image information is displayed in a Ganzfeld (Gur, 1989, 1991). It is thought that successive contour transitions across luminance boundaries maintain transient responses from receptive fields but also stabilize activity of "luxotonic" cells that register brightness and color (Gur, 1987; Gur and Snodderly, personal communication). We submit that the flicker-fused stimulus is providing temporal activation that is very similar to that generated by tremor and microsaccades. These movements provide successive glimpses of a given image location, similar to the successive flashes of a flicker-fused stimulus. We are proposing that being able to register the average intensity of a flicker-fused display derives from mechanisms that evolved to provide stable perception of contrast and luminance.

## Coda

A key feature of the Talbot-Plateau law is the requirement for reciprocity of flash duration and frequency. This principle often holds for photochemical reactions (Bunsen-Roscoe law) as well as photoelectric effect with purified silicon. But it seemed improbable that signaling systems of the visual system could average the intensity of a sequence of light flashes as a match to steady intensity, and do so for various combinations of flash intensity and duration. One might think this could be accomplished by persistence of stimulus influence for a 50% duty cycle, where the flash sequence provides light for half of each period and the average intensity is simply double the intensity of the steady source. But for an extreme combination, such as the one microsecond flash at 25 hertz, the period of darkness is essentially 40 ms. The energy of the flash would have to be measured with great precision and then drained out, like a leaky capacitor, at exactly the right rate to fill that void. It does not seem plausible that the reciprocity proscribed by the Talbot-Plateau law could be accomplished by visible persistence.

The present results indicate that the predictions of the Talbot-Plateau law are relatively precise across about seven orders of light intensity. Physiological evidence suggests that this is being accomplished by mechanisms that counterbalance signals from the ON and OFF channels, wherein the degree of departure from an average light level determines the degree of activation of each channel.

## Author contributions

EG: designed study, oversaw data collection, curated data, and wrote much of the manuscript JM: constructed equipment, programmed experiments, consulted on experimental plan, wrote portions of methods section, and provided manuscript edits. All authors contributed to the article and approved the submitted version.

## Acknowledgments

David Nyberg designed the circuitry for the display system. Respondents were tested by Bo Yan Moran, Hannah Nordberg, and Alyssa Brill. Statistical analysis and modeling was provided by Wei Wang, Department of Medicine, Harvard Medical School and Brigham and Women's Hospital. This research was supported by the Neuropsychology Foundation and the Quest for Truth Foundation.

## Conflict of interest

The authors declare that the research was conducted in the absence of any commercial or financial relationships that could be construed as a potential conflict of interest.

## Publisher's note

All claims expressed in this article are solely those of the authors and do not necessarily represent those of their affiliated organizations, or those of the publisher, the editors and the reviewers. Any product that may be evaluated in this article, or claim that may be made by its manufacturer, is not guaranteed or endorsed by the publisher.

## Supplementary material

The Supplementary material for this article can be found online at: https://www.frontiersin.org/articles/10.3389/fnins.2023.1169162/full#supplementary-material